\documentclass[showkeys,twocolumn,citeautoscript,showpacs,amsmath,amssymb,epsf,prl]{revtex4}
\usepackage{dcolumn}
\usepackage{epsfig}

\usepackage{graphicx}
\usepackage{dcolumn}
\usepackage{bm}

\def\etal{{\em et al. }}

\def\cm{ cm$^{-1}$ }
\def\AAA{\AA$^3$}

\def\cmin{ \cm$^{-1}$}

\begin{document}
\preprint{UTEP/005}

\title{ The vibrational stability and  electronic structure of B$_{80}$ fullerene-like cage}

\author{Tunna Baruah$^1$, Mark R. Pederson$^2$, , and
Rajendra R. Zope$^{1,3}$} 

\affiliation{$^1$Department of Physics, The University of Texas at El Paso, El Paso, TX 79958, USA }

\affiliation{$^2$Center for Computational Materials Science,  US Naval Research Laboratory, Washington, DC 20375, USA}

\affiliation{$^3$NSF CREST Center for Nanomaterials Characterization Science and Process Technology \\
    Howard University, School of Engineering, 2300 Sixth Street, N.W.  Washington, D.C. 20059, USA}

\date{\today }

\begin{abstract}

   We investigate the vibrational stability and  the electronic structure of the  proposed icosahedral 
fullerene-like cage structure of B$_{80}$ [Szwacki, Sadrzadeh, and Yakobson, Phys. Rev. Lett. {\bf 98}, 166804 (2007)] by an all
electron density functional theory using polarized Gaussian basis functions containing
41 basis functions per atom. The vibrational analysis of B$_{80}$ indicates that the icosahedral structure
is vibrationally unstable with 7 imaginary frequencies.  The equilibrium structure has  $T_h$  symmetry
and  a {\em  smaller} gap of 0.96 eV between the highest occupied and lowest unoccupied 
molecular orbital energy levels compared to the icosahedral structure.  The static dipole polarizability 
of B$_{80}$ cage is 149 \AAA and the first ionization energy is 6.4 eV. The B$_{80}$ cage has rather large 
electron affinity of 3 eV making it useful candidate as electron acceptor if it is synthesized.
The infra-red and Raman spectra of the highly symmetric structure  are characterized by a few absorption  peaks.
\end{abstract}

\maketitle

\begin{figure}
\epsfig{file=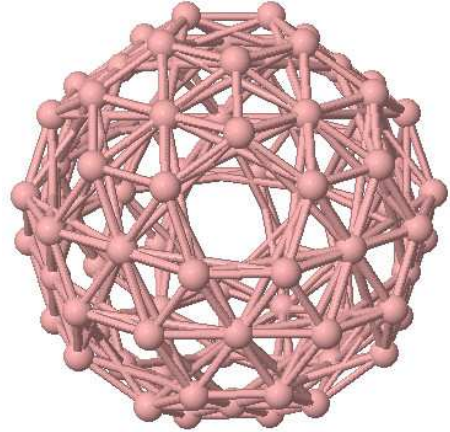,width=6.0cm,clip=true}
\caption{\label{fig:str} (Color online) Optimized geometry 
of  B$_{80}$ cage.
}
\label{fig1}
\end{figure}

  Since the discovery of C$_{60}$, several studies have reported possible 
existence of hollow inorganic cage-like structures. Recent work by Szwacki etal\cite{RF:455}
added boron to the list of elements that can form fullerene-like 
hollow cage structure. While several 
studies\cite{RF:456,RF:459,RF:458,RF:457,RF:454,PhysRevB.68.035414,lau:212111,boustani:3176,BQ97}
have reported stable clusters, rings, and nanotubes of boron, a hollow fullerene-like
cage cluster containing only boron has not yet been found.  
Using density functional 
theory Szwacki etal\cite{RF:455}, showed that a boron cluster containing 80 atoms can 
form a stable  hollow cage.  The basic structure of this cluster is similar 
to that of C$_{60}$ with  12 pentagonal and 20 hexagonal rings. The B$_{80}$ 
cluster has an additional boron atom at the center of each hexagon (Cf. \ref{fig1}). 
It is known that the boron analogue of C$_{60}$ is not a stable structure. However, these  additional 20 
boron atoms at the centers of the hexagonal rings  stabilize the B$_{60}$ to form a 
stable B$_{80}$ fullerene with a binding 
energy of 5.77 eV per atom.
The structure of B$_{80}$ cluster has been suggesteded by Szwacki etal.
to maintain  the same 
icosahedral point group symmetry as in C$_{60}$ fullerene. They also predicted that  the  
icosahedral B$_{80}$, as in the case of C$_{60}$, has the lowest unoccupied molecular orbital (LUMO) of 
t$_{1u}$ symmetry.  However they find that the highest molecular orbital (HOMO)  however belongs to the 
t$_{2u}$ irreducible representation\cite{RF:455}. 

\begin{figure}
\epsfig{file=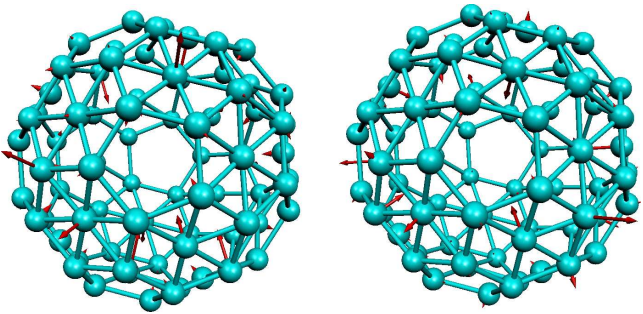,width=8.5cm,clip=true}
\caption{ (Color online) Unstable 
vibrational modes  of icosahedral B$_{80}$ cage.}
\label{fvib2}
\end{figure}

   The purpose of the present article is to further investigate the electronic 
structure of this novel boron cluster, focusing  particularly on its 
response to applied electric field. To this end, we calculate the dipole 
polarizability of the B$_{80}$ cluster using the linear combination of 
atomic orbital approach within the density functional formalism. In 
addition to the dipole polarizability we also compute the first ionization 
energies and the electron affinity. We also provide the infra-red and Raman 
spectra for possible detection. We use large polarized Gaussian 
basis sets  to express the Kohn-Sham 
molecular orbitals\cite{RF:181}. The exchange-correlation effects are treated 
within the generalized gradient approximation using the 
Perdew-Burke-Ernzerhof\cite{RF:183} parametrization. We have first 
optimized the structure of B$_{80}$ using the icosahedral point group 
by the NRLMOL code\cite{RF:69,RF:185}.  The resultant electronic 
structure shows some discrepancy with respect to the
previous study by Szwacki\cite{RF:455}. The HOMO level is 
5 fold degenerate and belongs to the h$_u$ irreducible representation 
instead of t$_{2u}$ as reported earlier. 
Besides, this small discrepancy, the present calculation also revealed that a more stable B$_{80}$
cluster has symmetry lower than the icosahedral symmetry as reported by Swazcki  
\etal. To verify if this lower symmetry structure is due to symmetry 
breaking distortion of icosahedral cage, full vibrational frequency 
calculations were performed within the harmonic approximation.
These calculations show that the icosahedral B$_{80}$ cluster 
is vibrationally unstable with two sets of imaginary frequencies. 
One of the two sets of modes is four fold degenerate and 
the the other one is a three fold degenerate mode. The vibrational
motions of atoms in these two modes are depicted in Fig. \ref{fvib2}. These modes 
are the out of plane motion of the boron atoms capping hexagons.

Full symmetry unconstrained relaxation of B$_{80}$ cluster show that
the cluster  has $T_h$ symmetry.  The $T_h$ cluster is vibrationally 
stable.
The $T_h$ B$_{80}$ cluster has five inequivalent atoms whose 
positions in Bohr are 
      (0.0000,     7.9117,     1.6024),
      (4.3303,     4.3303,     4.3303), 
      (2.6304,     6.8420,     3.2061),
      (5.1888,    -5.8587,    -1.6209), and
      (0.0000,     2.5093,     6.6176).
The position of all atoms in B$_{80}$ cluster can be obtained using the 
position of the nonequivalent atoms and using the symmetry 
operations of $T_h$. The 4 three fold symmetry axes are along 
the 111 directions.

 The energy of T$_h$ structure is 
lower by 0.05 eV compared to the icosahedral structure.
Its electronic structure is 
7a$_{u}$ 12a$_{g}$ 22e$_{g}$ 22t$_{g}$ 27t$_{u}$ 12e$_{u}$.
The h$_u$ symmetry of the HOMO of the icosahedral structure is split into a three-fold
degenerate t$_u$ and a  two fold degenerate e$_u$ level in the $T_h$ structure. The 
splitting also reduces the HOMO-LUMO gap in the $T_h$ structure.
The HOMO is two fold degenerate and belongs to $e_u$ irreducible
representation. The LUMO is three fold degenerate is of $t_u$
type.  
The HOMO and LUMO orbital densities are shown in Fig.\ref{MO:fig}.

    The static dipole polarizability is an important 
physical property that characterizes the system's response 
to an applied static electric field to the first order. A number of 
methods have been developed to compute the dipole and higher 
polarizabilities. In this work, we use the finite-field 
method. In the finite-field method the total energies and/or dipole moments 
are computed self-consistently  for various values of the applied electric 
field\cite{cohen:s34,Bartlett_FF,Kurtz82,Salahub_Pol,Na_Pol,Jackson}. 
The polarizability tensor is then built from well converged 
total energies or dipole moments using the finite-difference 
approximation. An accurate estimate of the polarizability 
using the linear combination of atomic orbitals requires the use of a large basis set
supplemented with  diffuse functions. The Gaussian basis 
set used in this work consists of 5 $s-$, 4 $p-$, and 3 $d-$ type Gaussians each 
contracted from 12 primitive functions. This basis is augmented with 6 
$d-$ type functions. Thus, in total 3280 basis functions are 
used in polarizability calculation. More details about construction
and the performance of basis set can be found 
in Ref. \onlinecite{RF:181,PBAS05,PB_POL}.
Due to the quasi spherical symmetry of B$_{80}$ cluster, the off-diagonal 
elements of the polarizability tensor are zero. The mean polarizability 
obtained by finite-field method is 149 \AAA. The mean 
polarizability of C$_{60}$ fullerene determined using the same set 
of approximations is 82 \AAA. The larger polarizability 
of B$_{80}$ cluster is principally due to 
its larger volume.  Using classical electrostatics, the polarizability of 
a spherical shell of radius R can be shown to be R$^3$. Unlike the
C$_{60}$ fullerene where in all atoms are at same distance 
from its center of mass, the atoms in B$_{80}$ cluster are at slightly 
different distances (3.74-4.27\AA) from the center of B$_{80}$ cluster.
Using the radius of the outermost atoms, the volume of B$_{80}$ cluster can be estimated to 
be roughly 1.74 times that of the C$_{60}$ fullerene. Using 1.74 as a scaling factor, the 
polarizability of C$_{60}$ fullerene can be used to estimate polarizability of B$_{80}$ 
cluster. This rough estimate is 143\AAA, in good agreement with the polarizability
obtained by finite-field method. 
%
%
 The finite-field (screened) polarizability does 
not change due to  symmetry lowering of the B$_{80}$ cluster. The unscreend 
polarizability however show significant change upon symmetry lowering. 
It increases from 631\AAA to 897 \AAA due to the decrease in the HOMO-LUMO 
gap and due to changes in low-lying dipole allowed transitions. These 
transitions must be strongly screened to give identical values 
of polarizability for $I_h$ and $T_h$ clusters. 

\begin{figure}
\epsfig{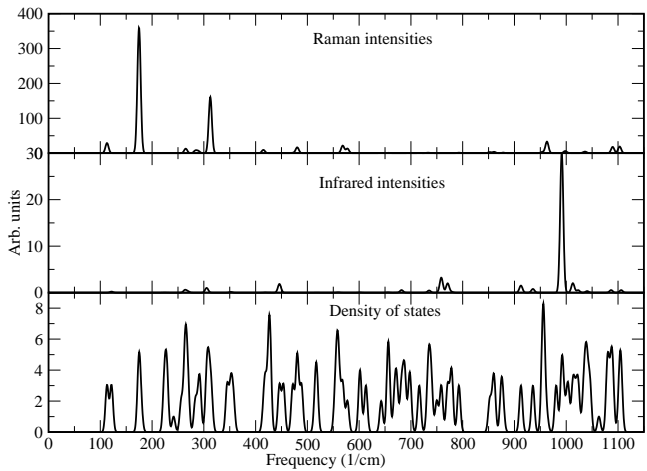}
\caption{ (Color online) The vibrational density of states, Infrared absorption 
intensity and the Raman scattering intensities of the B$_{80}$ cluster with T$_h$ structure.
\label{vib}}
\end{figure}

\begin{figure}[b]
\epsfig{file=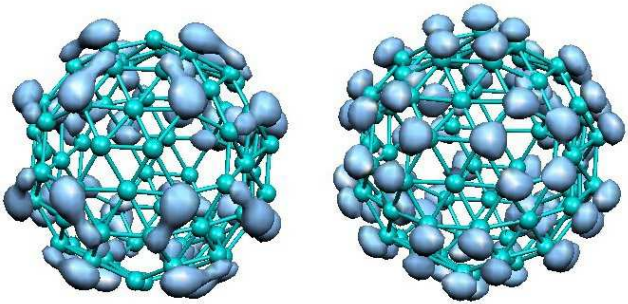,width=8.5cm,clip=true}
\caption{ (Color online) The orbital densities 
of the HOMO (left) and LUMO (right) of the B$_{80}$ cluster.
\label{MO:fig}}
\end{figure}

The calculation of the vibrational  frequencies establishes the stability 
of the T$_h$ structure.
 The vibrational density of states of the B$_{80}$ cluster are shown in 
fig. \ref{vib}. 
 The bottom panel shows the density of states and the upper two panels 
show respectively the infrared and Raman activity of the B$_{80}$ cluster. The cluster
shows a very strong infrared peak at 991 cm$^{-1}$. The other significant peaks
occur at 446, 759, 771, 912, and 1012 cm$^{-1}$. The Raman spectrum shows a few
low frequency modes with strong peaks at 112, 174, 176, and 312 \cmin.
Another prominent peak is seen at 963 \cmin.

The vibrational frequencies
determined in analysis of the vibrational stability of B$_{80}$ cluster can be
also be used to compute the vibrational contribution to the dipole 
polarizability. 
In general, the vibrational polarizability  is the second largest contribution 
to the total polarizability tensor.  
 For the  case of ionic and hydrogen bonded systems, the vibrational contribution 
may be comparable to or even larger than the electronic contribution.
The vibrational contribution to the polarizability tensor within 
the double harmonic approximation\cite{PBAS05} is given as 
$$\alpha^{vib}_{i,j} = \sum_{\mu} Z_{i,\mu} \omega_{\mu}^{-2} Z_{j,\mu}^T.$$
Here, $\omega_{\mu}^{-2}$ is the frequency of the $\mu$th vibrational mode, 
$ Z_{i,\mu}$ is the effective charge tensor (See. Ref. \onlinecite{PBAS05} for more details).
The vibrational contribution to polarizability is 5.5 \AAA which 
is much smaller the electronic contribution but  is 
larger than that observed in carbon fullerenes\cite{PBAS05}.

  To understand  the reactivity of the B$_{80}$ cluster, we have calculated its 
ionization potential and the electron affinity. The first ionization energy 
is the minimum energy required to remove an electron.  It can 
be determined from the self-consistent total 
energy difference of the B$_{80}$ cluster and its singly charged cation\cite{RF:321}. The 
calculated ionization energy is 6.5 eV and is smaller than that of 
C$_{60}$ (7.6eV). Similarly, the electron affinity can be computed from 
self-consistent total energy difference of neutral cluster and its anion. 
The calculated electron affinity is 3 eV. Our calculations also indicated that
the cluster reorganization upon addition of an electron is also small 
in this highly symmetric cluster. 
The chemical hardness is an indicator of the reactivity of the molecule.
It can be approximated as half of the difference between the 
ionization potential and electron affinity\cite{hardness}. This definition immediately 
points out that the B$_{80}$ cluster is more reactive than the 
C$_{60}$ fullerene.
The rather large electron 
affinity makes B$_{80}$ cluster, if synthesized, an interesting candidate 
as an electron receptor. The larger electron affinity also suggests 
possibility of  coating these cluster with alkali or transition atoms, which 
then can be tested as hydrogen storage materials. Other possibilities are
using it as a building block in ionic cluster assembled materials or as an
electron receptor in a photovoltaic device.

 To summarize, the vibrational stability of recently reported B$_{80}$ 
cluster is examined by computing the harmonic vibrational frequencies.  
The B$_{80}$ cluster is found to be unstable in icosahedral symmetry but stable in 
the reduced $T_h$ symmetry. The electronic structure of B$_{80}$ cluster changes upon 
symmetry lowering. The symmetry lowering decreases the HOMO-LUMO gap from 1.1 eV (in I$_h$ structure) 
to 0.97 eV.
Its static dipole polarizability (149\AAA) however does not alter appreciably. 
Its ionization potential is 6.5 eV. The $T_h$ structure 
has large electron affinity of 3 eV making it a candidate as an 
electron receptor. Using the chemical hardness as an indicator 
of reactivity the  B$_{80}$ is found to be larger than 
the C$_{60}$ fullerene. The infra-red and Raman spectra are provided.

        This work is supported in part by 
the National Science Foundation through CREST grant,
by the University of Texas at El Paso (UTEP startup funds, University research institute grant) and  partly
the Office of Naval Research, directly and through the Naval Research Laboratory. 
Authors acknowledge the computer time at the UTEP Cray 
acquired using ONR 05PR07548-00 grant.


\end{document}